# PHASE TRANSITIONS IN BINARY CATEGORIZATION: EVIDENCE FOR DUAL-SYSTEM DECISION MAKING

Ihor Lubashevsky, Ian Wilson, Rio Suzuki, Yuki Kato
*University of Aizu, Ikki-machi, Aizu-Wakamatsu, Fukushima, 965-8580 Japan*
<IL: i-lubash@u-aizu.ac.jp, IW: wilson@u-aizu.ac.jp>

## Abstract

*We report experiment results on binary categorization of (i) gray color, (ii) speech sounds, and (iii) number discrimination. Data analysis is based on constructing psychometric functions and focusing on asymptotics. We discuss the transitions between two types of subjects' response to stimuli presented for two-category classification, e.g., visualized shade of gray into "light-gray" or "dark-gray." Response types are (i) the conscious choice of non-dominant category, described by the deep tails of psychometric function, and (ii) subjects' physical errors in recording decisions in cases where the category choice is obvious. Explanation of results is based on the concept of dual-system decision making. When the choice is obvious, System 1 (fast and automatic) determines subjects' actions, with higher probability of physical errors than when subjects' decision-making is based on slow, deliberate analysis (System 2). Results provide possible evidence for hotly debated dual-system theories of cognitive phenomena.*

## Introduction

In the present work we report the results of our experiments on binary categorization of stimuli from two different sensory modalities (auditory and visual) and objects (numbers) whose perception is pure mental. Conducting these experiments we pursued two goals:

- To find evidence for the universal mechanism of decision making in categorization which assumes that
  - the corresponding sense organ just converts an external stimulus into a neurophysical signal bearing the information about the stimulus proximity to the analyzed categories encoded in some general manner;
  - the brain processes this signal in a way independent of particular details characterizing the involved sensory modality.
- To verify whether it is possible to discriminate between situations when the decision-making in categorization is governed solely by physiological mechanisms and when deliberate analysis contributes substantially to categorization.

The reasons for posing these issues are as follows. First, the data accumulated in fMRI, EEG, and neuropsychological investigations enabled Walsh and Bueti (Walsh 2003, Bueti and Walsh 2009) put forward a new paradigm about human judgment and evaluation of external stimuli called ATOM ("A Theory Of Magnitude"). The ATOM supposes that various dimensions of magnitude information are encoded by "common neural metrics" in the parietal cortex, which explains the emergence of common neurocognitive mechanisms governing human perception of various physical stimuli. Hayes et al. (2014) generalized the ATOM by extending its scope onto memory, reasoning, and categorization. traditionally treated as separate components of human cognition. The exemplar-based account of the relationship between categorization and recognition (Nosofsky et al. 2012) is also rather close to this paradigm.

Second, as demonstrated (Baird and Noma 1975, Noma and Baird 1975) human perception of physical stimuli based on our sense organs and the mental evaluation of abstract objects like numbers are similar in the basic properties.

Third, nowadays the dual-processing account of human behavior is widely used in cognitive psychology. It holds that there are two distinct processing systems available for cognitive tasks. System 1 is fast, automatic and non-conscious, System 2 is slow, controlled and conscious. Whether the two systems do exist at the level of neurological processes or they admit the interpretation as individual subsystems with own properties is a subject of on-going debates, for arguments for and against a reader may be refereed to Rustichini (2008), Evans (2008, 2011), Kahneman (2011), Barrouillet (2011). Moreover, nowadays the idea about the cumulative contribution of two different mechanisms—fast guesses and slow controlled decisions–to the speed-accuracy tradeoff (Ollman 1966) has become popular; for a review see, e.g., Heitz (2014).

The purpose of our experiments was to accumulate enough statistical data to analyze the *asymptotics* of the corresponding psychometric functions accompanied with the dependence of the mean decision time on the uncertainty in category choice. As we demonstrated previously (Lubashevsky and Watanabe 2016, Namae et al. 2017), the asymptotics of psychometric function bears the information enabling one to discriminate between plausible mechanisms governing the categorization process in a clear way.

### Gray color categorization

Each trial of color categorization was implemented as follows. A random integer $I \in [0, 255]$ is generated and some area on PC monitor is filled with the gray color $G(I) := RGB(I, I, I)$. Then a subject has to classify the visualized gray shade $G(I)$ according to his/her perception into two possible categories, "light gray" and "dark gray." A made choice is recorded via pressing one of two joystick buttons. Then a mosaic pattern of various shades of gray is visualized for 500 ms to depress a possible interference between color perception in successive trials that can be caused by human iconic memory. After that a new number $I$ is generated and the next trial starts. The moment when a subject presses the button are also recorded, which gives the decision time in the current trial.

The experiments were set up as follows. Four subjects, two female and two male students of age 21–22 were involved in these experiments. The experiments spanned 5 successive days, for each subject the total number of data records was 2,000 data-points per day and finally 10,000 for 5 days. One day set comprised four blocks of 15 min experiments separated by 3 min rest. For each subject the total data-set aggregates all the records collected during 5 days. No special instructions were given to the subjects about the necessity to make decision in selecting categories as quickly as possible.

Figure 1 illustrates the obtained results which allow us to draw the following conclusions.

1. In the log-normal scales the asymptotics of psychometric functions for both the categories can be approximated by a linear dependence on the number $I$, which is the characteristic feature of potential mechanism (Lubashevsky and Watanabe 2016). It means that the decision-making may be regarded as a probabilistic event of finding a particle in an equilibrium state described by some potential. The region of this asymptotic behavior matches the most pronounced uncertainty in selecting the categories.
2. As the analyzed gray shade (the number $I$) penetrates deeper in the region of rare events (rare choice of inappropriate category) this linear asymptotics is replaced by the probability of choosing the inappropriate category that does not decrease or even can increase with the further change in the number. The transition between these modes of the choice probability behavior is well pronounced, which is noted by arrows in Fig. 1.
3. The increase of the mean decision time in the region of choice uncertainty typically exceeds 1 s, which is substantially longer than the upper boundary of human response delay controlled by pure physiologically processes. It argues for a significant contribution of mental processes to the categorization in the region of its essential uncertainty.

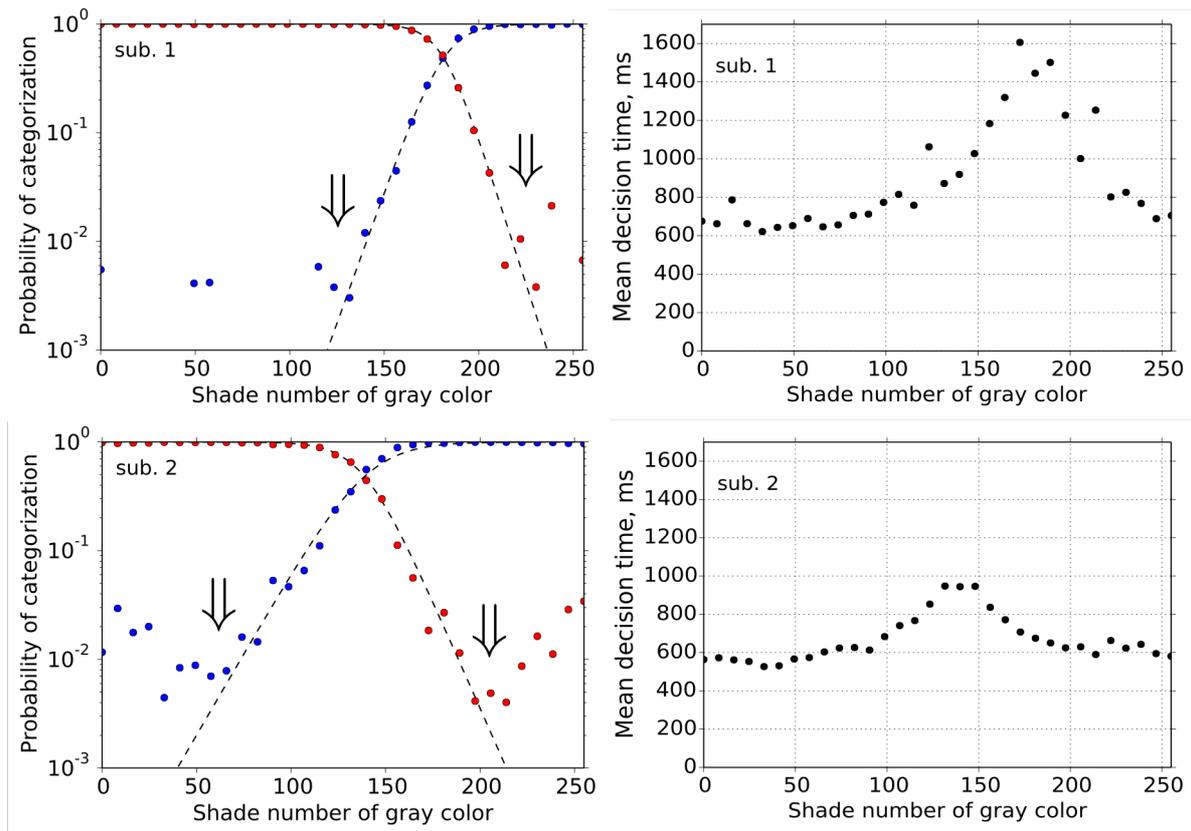

Fig. 1. *Left column:* Psychometric functions of gray color categorization, data points corresponding to the "light-gray" and "dark-gray" classes are shown in blue and red, respectively. Dashed lines represent fitting functions of logistic form. *Right column:* Mean decision time in gray color categorization depending on the shade number of gray.

### Vowel sound categorization

Using the "Create Sound from Vowel Editor…" menu item in Praat acoustic analysis software (Boersma & Weenink, 2018), we synthesized a total of 68 sound files. Each sound was a 100-millisecond vowel with a different second-formant (F2) frequency, which ranged from 810 Hz ([o]) to 2,150 Hz ([e]) in 20 Hz steps. The first-formant (F1) was kept constant at 400 Hz for all sounds. So, the first sound's (F1, F2) values in Hz were (400, 810); the second sound's values were (400, 830), the third sound's values were (400, 850), and so on up to the 68th sound, which had values (400, 2150). Thus, the 68 sounds varied along a single dimension – that of F2. The pitch of all sounds was kept constant at 140 Hz–a pitch within the normal speaking range of a human voice. The 68 unique audio files were randomly ordered to make a "cycle", and this was done 12 times to make 12 different cycles. A randomly ordered group of 12 cycles was considered 1 trial and contained 68 x 12 = 816 vowel sounds. Each subject completed 10 trials (each with a random order of cycles) with a short break between each one, for a total of 8,160 vowel sounds.

Two healthy subjects with no reported hearing problems participated in this research. They were both Japanese 4[th]-year undergraduate students who were in their early twenties. Subject 1 was female and Subject 2 was male. E-prime 2.0 software (*Psychology Software Tools*) and a 5-button "Chronos" multifunctional response and stimulus device were used for recording subject actions. The Chronos device has millisecond accuracy and consistent sound output latencies across machines. Subjects listened to the stimuli through JVC headphones and were instructed to react as quickly as possible to identify the stimuli that they heard.

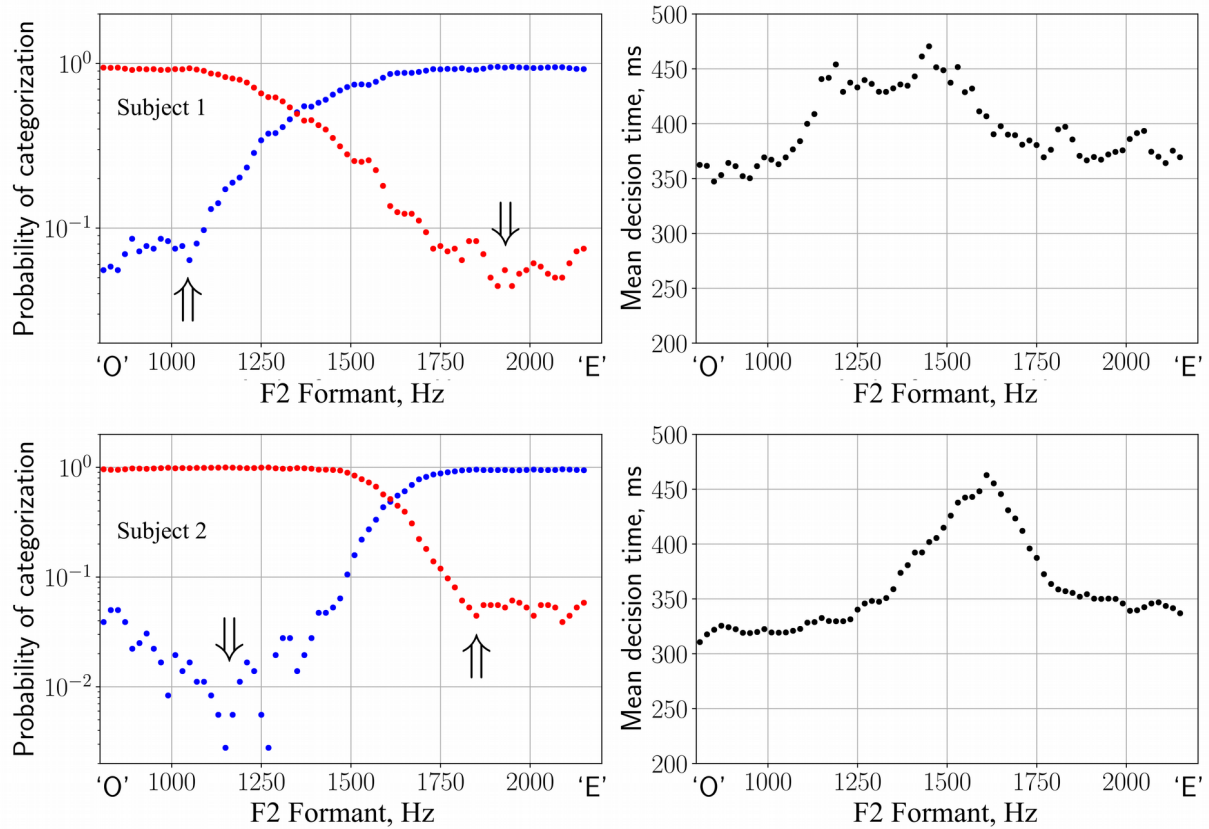

Fig. 2. *Left column:* Psychometric functions of vowel sound categorization, data points corresponding to [o] and [e] classes are shown in red and blue, respectively. *Right column:* Mean decision time in vowel sound categorization depending on the F2 Formant.

Figure 2 exhibits the obtained results which also argue for conclusions 1 and 2 stated in the previous section for the color categorization. The main difference is that the mean decision time in its maximum attains the upper boundary of the response delay time usually attributed to the information processing by neural networks. However, the similarity of the asymptotic behavior of psychometric functions in both the cases allows us to hypothesize that mental processes play a significant role also in the sound categorization in the region of its substantial uncertainty. The reaction of subjects instructed to respond as quickly as possible just corresponds to the minimal delay of the conscious response coinciding with the maximum of physiological delay such that no gap in the human response delay time appears.

## Categorization of numbers

The idea of the experimental setup of studying the binary categorization of numbers is similar to that of color categorization. An integer I from the interval [1, 256] is randomly selected and visualized. A subject has to select or reject a visualized integer, which is recorded via pressing the corresponding button of joystick. The conditions of experiments stimulated subjects to select integers near the right boundary of the interval [1, 256] and to reject integers near its left boundary with certainty. The uncertainty of this choice becomes significant for intermediate integers and the width of the corresponding region is estimated as 50. Four subjects—male students of about 20 years old—were involved in these experiments. The experiments were done for 4 sets with about 2,000 data points per set; the total amount of recorded data for each participant is about 8,000 data-points.

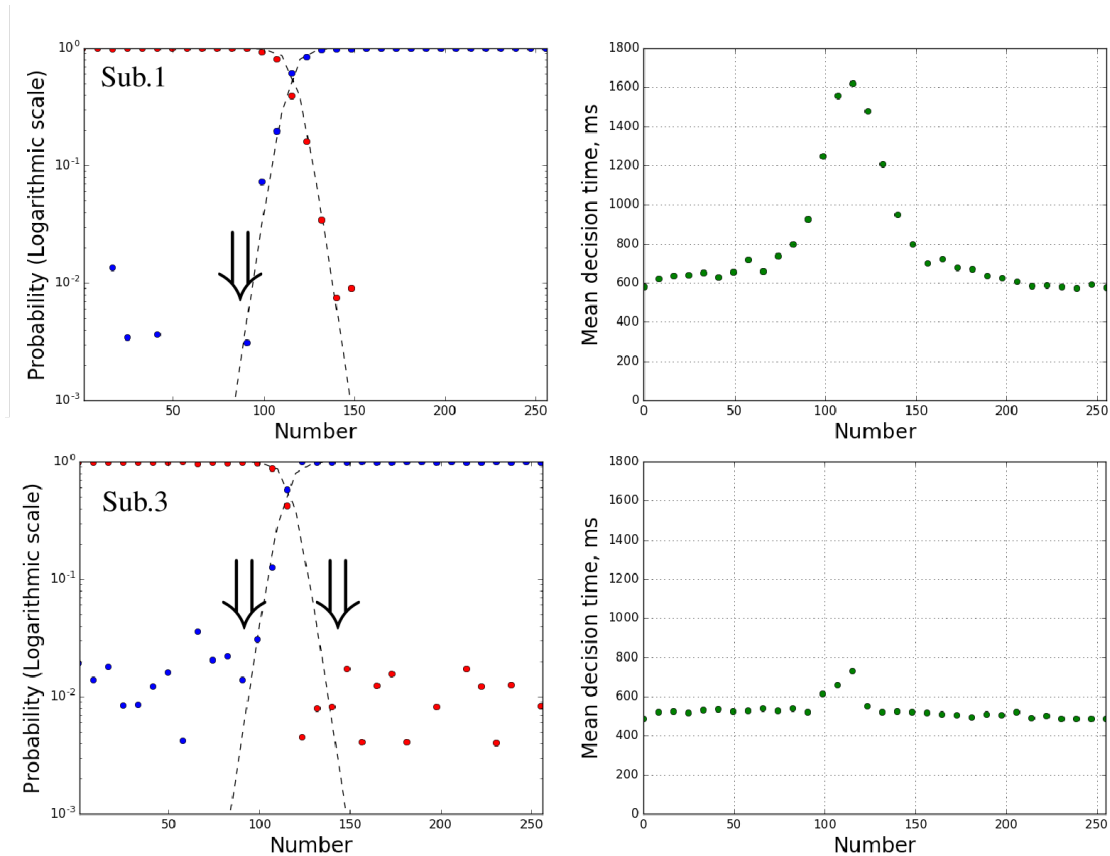

Fig. 3. *Left column:* Psychometric functions of number categorization, data points corresponding to "rejecting" and "accepting" a given number are shown in red and blue, respectively. Dashed lines represent fitting functions of logistic form. *Right column:* Mean decision time in number categorization (based on Nihei, 2018).

Figure 3 illustrates the obtained results. The categorization of numbers under the given conditions may be treated as some mixture of the color and sound categorization in property.

## Conclusion

The presented results enable us to posit the following.
- The choice between the auditory and visual categories as well as abstract categories in number comparison is governed by a universal central mechanism of the potential type. It is reflected in the linear asymptotics of psychometric functions in log-normal scales, which corresponds to significant uncertainty in categorization.
- There is a sharp transition between the linear asymptotic behavior of psychometric functions and their behavior when the choice of appropriate category is obvious. We relate the linear asymptotic behavior to conscious choice governed by slow System 2; it is characterized by the regular growth of decision time as the choice uncertainty increases. The choice of appropriate category when it is obvious seems to be governed by automatic, fast System 1. It explains the relative increase in the probability of choosing the inappropriate category by human errors in motor behavior when the conscious control is depressed.
- When the choice uncertainty is high, mental processes seem to affect substantially the decision time and its maximum can be used to quantify the relative contribution of conscious and unconscious processes as a whole. In particular, the shorter the maximal decision time, the hight the contribution of unconscious component, the higher the probability of motor errors in the case of System 1 control.

## Acknowledgements

The authors wish to acknowledge the University of Aizu for Competitive Research Funding that was used in carrying out this research.

## References

Baird, J. C. and Noma, E. (1975). Psychophysical study of numbers: I. Generation of numerieal responses, *Psychological Research* **37**(4), 281–297.

Barrouillet, P. (2011). Dual-process theories and cognitive development: Advances and challenges, *Developmental Review* **31**(2–3), 79–85. Special Issue: Dual-Process Theories of Cognitive Development.

Boersma, P., & Weenink, D. (2018). Praat: doing phonetics by computer [Computer program]. Version 6.0.40, retrieved 11 May 2018 from http://www.praat.org/

Bueti, D. and Walsh, V. (2009). The parietal cortex and the representation of time, space, number and other magnitudes, *Philosophical Transactions of the Royal Society B: Biological Sciences* **364**(1525), 1831–1840.

Evans, J. S. (2008). Dual-processing accounts of reasoning, judgment, and social cognition, *Annual Review of Psychology* **59**, 255–278.

Evans, J. S. (2011), Dual-process theories of reasoning: Contemporary issues and developmental applications, *Developmental Review* **31**(2–3), 86–102. Special Issue: Dual-Process Theories of Cognitive Development.

Hayes, B. K., Heit, E. and Rotello, C. M. (2014). Memory, reasoning, and categorization: parallels and common mechanisms, *Frontiers in Psychology* 5, Article 529 (9 pages).

Heitz, R. P. (2014). The speed-accuracy tradeoff: history, physiology, methodology, and behavior, *Frontiers in Nuroscience*, **8**, Article 150 (19 pages).

Kahneman, D. (2011). *Thinking, Fast and Slow*. New York: Farrar, Straus and Giroux.

Lubashevsky I. and Watanabe, M. (2016). Statistical Properties of Gray Color Categorization: Asymptotics of Psychometric Function. *In Proceedings of the 47th ISCIE International Symposium on Stochastic Systems Theory and Its Applications, Honolulu, 2015*, Kyoto, Institute of Systems, Control and Information Engineers (ISCIE), pp. 41–49.

Namae, R., Watanabe, M. and Lubashevsky I. (2017). Gray Color Multi-Categorical Perception: Asymptotics of Psychometric Function. In *Proceedings of the 48th ISCIE International Symposium on Stochastic Systems Theory and Its Applications, Fukuoka 2016,* Kyoto, Institute of Systems, Control and Information Engineers (ISCIE), pp. 76–80.

Nihei, K. (2018). Statistical Properties of Number Classification: Asymptotics of Psychometric Function and Mean Decision Time, Thesis, University of Aizu, supervised by Lubashevsky I.

Noma, E. and Baird, J. C. (1975). Psychophysical study of numbers: II. Theoretical models of number generation, *Psychological Research* **38**(1), 81–95.

Nosofsky, R. M., Little, D. R. and James, T. W. (2012). Activation in the neural network responsible for categorization and recognition reflects parameter changes, *Proceedings of the National Academy of Sciences of the United States of America* **109**(1), 333–338.

Ollman, R. T. (1966). Fast guesses in choice reaction time, *Psychonomic Science* **6**(4), 155–156.

Rustichini, A. (2008). Dual or unitary system? Two alternative models of decision making, *Cognitive, Affective, & Behavioral Neuroscience* **8**(4), 355–362.

Walsh, V. (2003). A theory of magnitude: common cortical metrics of time, space and quantity, *Trends in Cognitive Sciences* **7**(11), 483–488.